\documentclass[aps,prb,twocolumn,amsmath,amssymb,nofootinbib,showpacs,superscriptaddress]{revtex4-2}

\usepackage[english]{babel}
\usepackage{latexsym}
\usepackage{graphics}
\usepackage{graphicx}
\usepackage{epsfig}
\usepackage{color}
\usepackage{bm}
\usepackage{amsmath}
\usepackage{amssymb}
\usepackage{amsthm}
\usepackage{dcolumn}
\usepackage{bbm}
\usepackage{float}
\usepackage{hyperref}
\usepackage{color}
\usepackage{epstopdf}
\usepackage{cleveref}
\usepackage[svgnames]{xcolor}
\usepackage{physics}
\usepackage[sort&compress]{natbib}
\usepackage{bbold}
\usepackage{enumerate}

\hypersetup{hidelinks,colorlinks=true,allcolors=DarkBlue}

\begin{document}
	%%%%%%%%%%%%%%%%%%%%%%%%%%%%%%%%%%%%%%%%%%
	\preprint{APS/123-QED}
	
	\title{Disorder-induced decoupling of attracting identical fermions: transfer matrix approach}
	\author{Lolita I. Knyazeva}
	\email{kniazeva.li@phystech.edu}
	\affiliation {Russian Quantum Center, Skolkovo, Moscow 143025, Russia}
	\affiliation {Moscow Institute of Physics and Technology, Dolgoprudny 141700, Russia}
	\author{Vladimir I. Yudson}
	\email{vyudson@hse.ru}
	\affiliation{Laboratory for Condensed Matter Physics, HSE University,  Moscow, 101000 Russia}
	\affiliation {Russian Quantum Center, Skolkovo, Moscow 143025, Russia}
	
	\date{\today}
	
	\begin{abstract}
	We consider a pair of identical fermions with a short-range attractive interaction on a finite lattice cluster in the presence of strong site disorder. This toy model imitates a low density regime of the strongly disordered Hubbard model. In contrast to spinful fermions, which can simultaneously occupy a site with a minimal energy and thus always form a bound state resistant to disorder, for the identical fermions the probability of pairing on neighboring sites depends on the relation between the interaction and the disorder. The complexity of `brute-force' calculations (both analytical and numerical) of this probability grows rapidly with the number of sites even for the simplest cluster geometry in the form of a closed chain. Remarkably, this problem is related to an old mathematical task of computing the volume of a polyhedron, known as NP-hard. However, we have found that the problem in the chain geometry can be exactly solved by the transfer matrix method. Using this approach we have calculated the pairing probability in the long chain for an arbitrary relation between the interaction and the disorder strengths and completely described the crossover between the regimes of coupled and separated fermions.
		
	\end{abstract}

	\maketitle
	
	\section{Introduction}\label{intro}
	
	The interplay of disorder and interaction is one of the central problems of condensed matter physics. It gives rise to a plethora of fundamental phenomena including metal-insulator and superconductor-insulator transitions, `bad metals', spin and electron glasses, to mention just few \cite{Mott2001, Altshuler1, Lee, Gantmakher, de2006random, Pollak}. Recently some of these traditionally condensed matter topics have also become a subject of interest in systems of ultracold atoms in optical traps \cite{Bloch, DeMarco}.
	
	A paradigmatic platform to study these phenomena in a quantum ensemble of interacting particles is the seminal Hubbard model \cite{Hubbard} with its very rich and complicated physics.
	In this Article we consider a simplified version of this model for identical fermions with a short-range attractive interaction in the limiting regime of low density and strong site disorder. Assuming the intersite hopping parameter to be small compared to both the interaction and the disorder we get a system where the quantumness of particles is suppressed (except for the fermionic statistics) and they may be considered as located on different lattice sites \cite{Anderson1958}. Finding the ground state and correlation functions for this kind of electron glass reduces to a statistical but still rather complicated problem. The particles can form clusters whose size distribution is determined by the relative strength of the interaction and the disorder.
	
	Here, as the first step towards the solution of the many-particle problem, we study
	a toy model of just two identical fermions restricted to a finite strongly disordered
	lattice with $N$ sites, so $2/N$ is an effective fermion `density' (filling factor).
	The quantity of interest is the probability $P_b$ of forming the `bound pair' of fermions
	due to their attractive nearest-neighbor interaction. In the case of non-identical fermions with an attractive on-site interaction, the model would be trivial: two fermions occupy a site with minimal energy and thus always form a bound state. But identical fermions should occupy different sites and if the disorder is
	stronger than the interaction, it may happen that the fermions located on neighboring sites
	have higher energy than those located on some separated (non-neighboring) sites.
	
	The probability to find two neighboring sites for which the energy of the attracting fermions is minimal, increases with the increase of the system size. Therefore, such particles
	in an infinite system necessarily form a bound state despite the presence of the disorder. However, in a finite size cluster the problem of energetically advantageous arrangement
	of two attracting fermions on neighboring or on distant sites becomes quite challenging.
	The probability $P_b$ should be calculated as a function of $N$ and of the interaction and the disorder strengths. Moreover, it depends on the connectivity
	of the lattice. For simplicity and having in mind the possibility of an exact analytical approach,
	we restrict our analysis to the one-dimensional case and consider a system in the form of a closed $N$-site chain.
	
	The condition that the bound state corresponds to the minimal energy, is represented by a
	set of ($\propto N^2$) linear inequalities for random potentials on the system sites,
	and the probability $P_b$ is determined by the averaging of these inequalities over the realization of the disorder. Remarkably, with the simplest - box-like distribution of the
	disorder this problem is equivalent to the calculation of the volume of
	a domain (polyhedron) in the $N$-cube restricted by a set of ($\propto N^2$) hypersurfaces. This
    problem is known to be NP-hard: owing to the rapidly (perhaps, faster than exponentially in $N$) growing complexity of the system, the possibility of straightforward brute-force calculations, both analytical and numerical, is practically restricted to small $N$.
	
	Nevertheless, somewhat surprisingly for a disordered
	system 	\cite{[A set of solvable models with interaction and disorder is scarce{;
			see, e.g.,} ]Derrida,*[]Derrida2}, it turns out that the
 	considered problem in the chain geometry allows for the transfer matrix approach.
	The eigenstates and eigenvalues of this transfer matrix obey an intricate integral equation.
    Another surprise is that this integral equation turns out to be solvable analytically. Implementing this approach we have calculated the pairing
	probability $P_b$ in the large $N$ limit for an arbitrary relation between the
	interaction and the disorder strengths.
	
	In section \ref{model} we describe the model and formulate a set of conditions imposed on the
	random potentials to yield the existence of the bound state. This section also illustrates
	the exact brute-force approach in the simplest nontrivial case -- the chain with $N=4$ sites,
	and describes problems of extending this approach to higher $N$. In section \ref{numerical} we present
	results of numerical (stochastic) experiments and quantitatively interprete them for the
	weak interaction case. In section \ref{TM} we develop the transfer matrix approach and calculate
	the probability $P_b$ in the large $N$ limit for an arbitrary relation between the interaction and the disorder strengths. In Conclusion we summarize the obtained results and discuss possible issues of further research.

	\section{The model}\label{model}

We consider a pair of identical fermions in a finite lattice cluster described
by the Hubbard model \cite{Hubbard} with the nearest-neighbor attractive interaction
$\widetilde{U}$. The short-range hopping
amplitude is assumed to be much smaller than both $\widetilde{U}$ and the disorder
distribution width $W$. Thus, the hopping has an almost negligible influence on the
ground state of the strongly disordered system and can be ignored. The aim of our
work is to find the probability $P_\text{b}$ that two fermions in the ground state
are `bound', i.e., they occupy two neighboring sites.
In an infinite cluster with the number of sites $N \rightarrow \infty$ the probability
$P_\text{b} \rightarrow 1$ for any nonzero $\widetilde{U}$, while in a finite cluster
it depends on relations between the parameters $\widetilde{U}$, $W$, and $N$, as well
as on the geometry (connectivity matrix) of the cluster. Calculation of $P_\text{b}$
as a function of  these parameters is a quite nontrivial problem. Here we restrict our
analysis to the simplest cluster in the form of a closed chain, where the solution can
be obtained by the transfer matrix method. For the random on-site potentials
$\widetilde{V}_i$ we choose the box probability distribution
$p(\widetilde{V}_i)= \theta(W - \widetilde{V}_i)\theta(\widetilde{V}_i )/W$,
though some of the derived expressions hold for a generic bell-shaped distribution.
The system Hamiltonian can be presented in the dimensionless form
\begin{equation}\label{H}
		H = \sum_{i=1}^{N} V_i n_i - U \sum_{i=1}^{N} n_i n_{i+1} \, ,
	\end{equation}
where $n_i$ is the number of fermions (0 or 1 fermion)
on the site $i$, and the total number of fermions on the chain equals two.
The interaction constant and the random potentials are measured in the units
of the disorder distribution width $W$: $U=\widetilde{U}/W$,
$V_i = \widetilde{V}_i/W$, so the disorder box-distribution function takes the form
\begin{equation}
		\label{eq:p_V}
		p(V_i) =\theta(V_i)\theta(1 - V_i) \, .
	\end{equation}
For a given disorder realization, the condition that the ground state of the
two-fermion system corresponds to the fermions located on some neighboring sites
(say, $i$ and $i+1$) means that the energy  $V_i + V_{i+1} -U$ is less than
energies of all other arrangements of fermions. If $U \geq 1$ this condition
of forming the bound state is fulfilled for any realization of disorder,
so $P_b(U \geq 1)=1$. Indeed, if the site $i$ corresponds to the minimal (in the
given realization) potential $V_i$, then the energy
$V_i + V_{i\pm 1} - U < V_j+ V_l$, for any pair $(j,l)$ of non-neighboring sites, as $V_i < V_j$ and  $V_{i\pm 1} - U < 0 < V_l$.

In the other limiting case, $U=0$, the non-interacting fermions occupy two
sites with minimal potentials. As the random potentials on different sites
are distributed independently, the probability for two fermions to occupy
neighboring sites is simply given by the ratio of the number $N$ of such
arrangements (in the considered ring geometry) to the total number
$C^2_N$ of possible arrangements:
\begin{equation}
		\label{pb0}
		P_b(U=0) = 2/(N-1) \, .
	\end{equation}
For brevity, we will refer to the fermion pair located on neighboring sites
as the `bound pair' even at $U = 0$ though the quantity (\ref{pb0}) is purely combinatoric.
Note that $P_b(U=0) \rightarrow 0$ when  $N \rightarrow \infty$.

Our task is to find $P_b(U)$ in the interval $0 < U < 1$. Due to the symmetry of the considered
ring cluster, the probability $P_\text{b}= N  P^{\text{b}}_{12}$, where $P^{\text{b}}_{12}$
is the probability
that the first and the second sites are occupied, i.e., the energy $V_1 + V_2 - U$ is lower
than energies of all other arrangements. This requirement is provided by a set of inequalities:
	\begin{eqnarray}
		V_1 + V_2 - U < V_j + V_l,\ j= \overline{2,N - 2},\ l= \overline{j+2,N};
		% \nonumber
         \label{eq:general_cond}\\
		V_2 - U < V_l,\ l=\overline{3,N-1};%\quad V_1 - U < V_l,\ l=\overline{4,N};
		%\nonumber
         \label{eq:particular_cond}\\
		V_1 + V_2 < V_j + V_{j+1},\ j=\overline{2,N-1};\quad
				V_2 < V_N,
		\label{eq:general_pairs}
	\end{eqnarray}
where the overlines indicate the intervals for site numbers $j$ and $l$. The inequalities (\ref{eq:general_cond}) and (\ref{eq:particular_cond})
mean that the energy of the fermion pair on the sites 1 and 2 is lower than that for fermions occupying  non-neighboring sites. 
The inequalities (\ref{eq:general_pairs})
ensure that the selected pair of sites (1,2) provides lower energy as compared to that for other
neighboring arrangements. The probability $P_b(U)$ is determined by the averaging of the above conditions
over the realizations of the random potentials:
	\begin{widetext}
		\begin{equation}
			\label{p12general}
			P_{b}(U) = N\bigg\langle \theta \left(  V_N - V_2  \right) \prod_{j=2}^{N-1} \theta \left( V_j + V_{j+1} - V_1 - V_2  \right)  \prod_{j=2}^{N-2} \prod_{l=j+2}^{N} \theta \left( V_j + V_l - V_1 - V_2 + U \right) \prod_{l=3}^{N-1} \theta \left(  V_l - V_2 + U \right)   \bigg\rangle \, .
		\end{equation}
	\end{widetext}
\noindent
Here $\langle \ldots \rangle $ means the averaging over the disorder. For the box distribution (\ref{eq:p_V}), this means the multiple integral over the unit N-dimensional hypercube.
The theta-functions in the integrand yield the conditions (\ref{eq:general_cond})-(\ref{eq:general_pairs}).

Being linear in random potentials, these conditions correspond to a set of hyperplanes restricting
the integration domain to a very intricate N-dimensional polyhedron. Finding the exact volume of a polyhedron is an old mathematical problem; computing this volume is NP-hard (see, e.g., \cite{Ong},
\cite{Bhatta-2023}, and references therein). Also, the direct analytical integration is highly repetitive
and tedious. For the particular integral (\ref{p12general}), \emph{a priori} we can say only  that it determines a polynomial function of the $N$-th order in $U$, that obeys (\ref{pb0}) and tends to unity when $U \rightarrow 1$. To illustrate the situation consider the simplest case $N=4$, where the integral (\ref{p12general}) can still be easily calculated.

\subsection{The simplest nontrivial case, $N=4$}\label{N4}
\begin{figure} [H]
		\centering
		\includegraphics[width=0.82\linewidth]{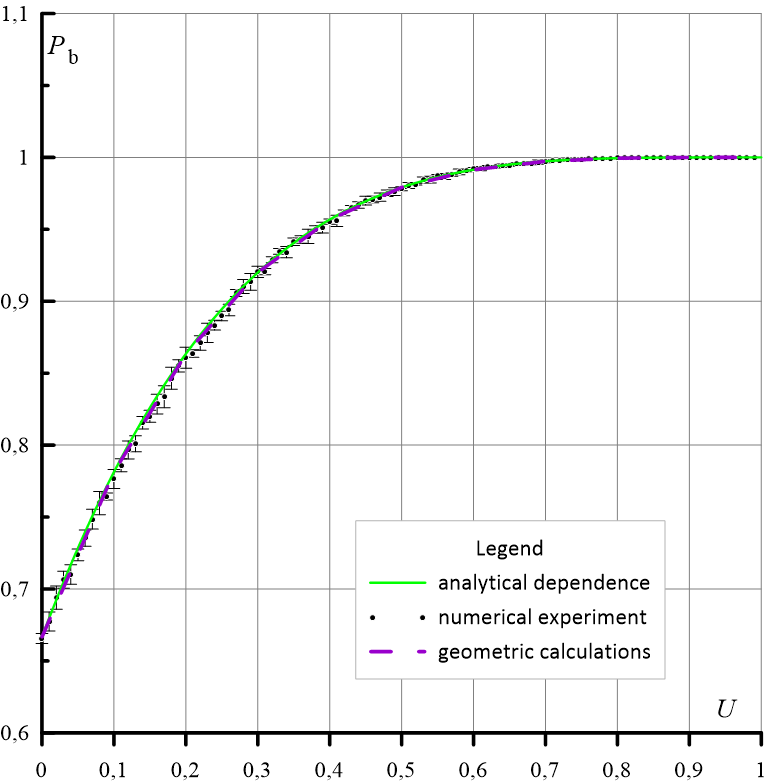}
		\caption{Dependence of the probability to observe a bound state $P_\text{b}$ on the interaction $U$ (in units of the disorder width) for the number of sites $N =4$.}
		\label{fig:n4}
	\end{figure}

The ring of ($N=4$) sites is the shortest one where not all sites are neighboring.
Using the conditions (\ref{eq:general_cond})-(\ref{eq:general_pairs}), we can represent
$P_{b}$ (\ref{p12general}) in the form
	\begin{equation}
		P_{b} = 4\int p(V_1) p(V_2) f\left( V_1, V_2\right)  f\left( V_2, V_1\right)  dV_1 dV_2,
	\end{equation}
	where
	\begin{eqnarray}
		f\left( V_1, V_2\right)=\int \theta \left( V_3 - V_2 + U\right) \theta \left( V_3 - V_1\right) p\left( V_3 \right) dV_3, \nonumber \\
		f\left( V_2, V_1\right)=\int \theta \left( V_4 - V_1 + U\right) \theta \left( V_4 - V_2\right) p\left( V_4 \right) dV_4. \,
	\end{eqnarray}
These expressions are valid for any distribution of random potentials. Substituting the uniform distribution (\ref{eq:p_V}), we obtain $P_\text{b}=4 P^{\text{b}}_{12}$:
	\begin{equation}
		\label{Pb_N4}
		P_\text{b} = 1 - \frac{1}{3} {\left( 1 - U \right)}^4.
	\end{equation}
\noindent In Fig. \ref{fig:n4} we see that this analytical dependence
coincides with the results of a direct numerical calculation of the integral as well as with
the numerical (Monte Carlo) experiment.
Even this simple example demonstrates the basic features of the effect: attracting fermions
on a finite lattice (ring) are coupled in the case of relatively weak disorder
$(U = \widetilde{U}/W > 1)$ but they can be decoupled in the case of relatively strong disorder $(U = \widetilde{U}/W < 1)$; if the disorder is very strong
$(U = \widetilde{U}/W \ll 1)$ the fermions are arranged almost independently.

\section{Numerical results and qualitative analysis}\label{numerical}
\begin{figure} []
	\centering
	\includegraphics[width=0.82\linewidth]{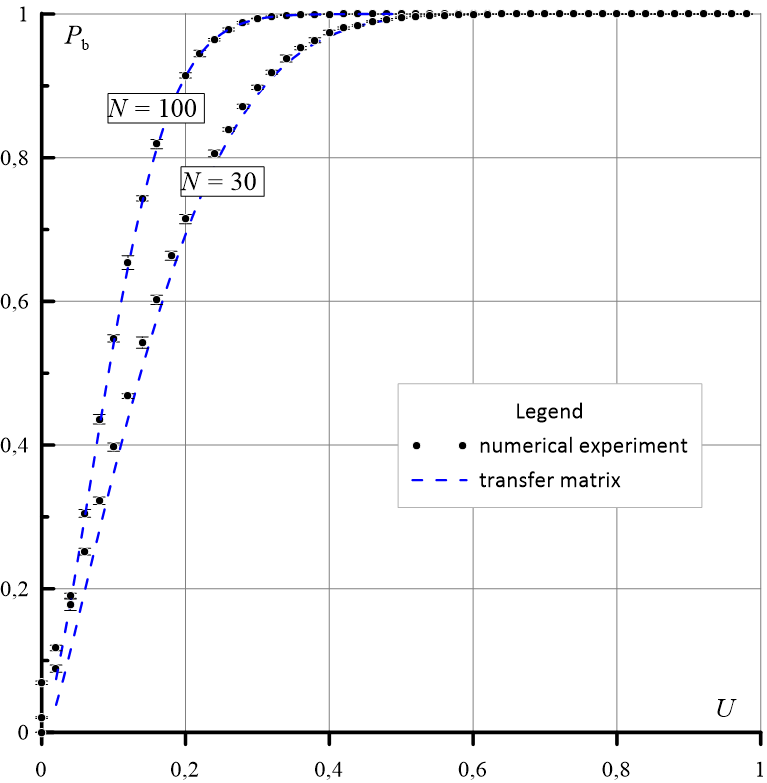}
	\caption{Crossover between the coupling and decoupling regimes. Dependencies of the probability $P_\text{b}$ on the interaction $U$ calculated in the numerical experiments for the number of sites $N=30$ and $N=100$ are presented.}
	\label{fig:crossover}
\end{figure}

The crossover between the coupling and decoupling
regimes depends on the cluster size (ring length
N), see dot results for the numerical (Monte Carlo) experiment in Fig. \ref{fig:crossover}. 
However, a description of these results by the brute-force analytical 
or numerical calculations can hardly be performed for a high $N$: with the increase of the number of sites $N$, the time
required for calculations grows very rapidly
(see Table \ref{tab:time}).
	\begin{table}[h]
		\caption{\label{tab:time}%
The time required to calculate the volume of a polyhedron in $N$-dimensional space that determines the probability of formation of a bound pair of identical fermions in a closed chain of $N$ sites. The results for the interaction value $U = 0.2$ are presented.
		}
		\begin{ruledtabular}
			\begin{tabular}{lllll}
				$N$& 4& 5& 6& 7\\
				\colrule
				 time, \text{s} & 3.407& 32.500& 632.812& 13273.593\\
			\end{tabular}
		\end{ruledtabular}
	\end{table}

To gain more insight into the problem, we will develop physical approaches and verify them with the results of numerical experiments. The numerical experiments have been performed as a search for the minimal energy in a particular realization of disorder. Counting the cases where the minimal energy corresponds to fermions occupying neighboring sites, we have calculated the probability of a bound state using 10000 realizations of the disorder.

We begin with the analysis of the weak interaction case $U \ll 1$.

\subsection{Weak interaction}\label{weak_interaction}
To warm up, consider a linear in $U \ll 1$ correction to $P_b(U=0)$, Eq.(\ref{pb0}).
First, let us show how the latter follows from the general expression (\ref{p12general}) due to the inequalities (\ref{eq:particular_cond}): in the non-interacting case they reduce to $V_l > \max{\{V_1, V_2\}}$,
for $l\in (3, ..., N)$, so the multiple integral (\ref{p12general}) takes the form
\begin{eqnarray}
		\label{pb-0-int}
		P_{b}(0) &=& 2N \int dV_1 p(V_1) \int dV_2 \theta(V_2-V_1) p(V_2) \nonumber \\
       & &\times  { \left[ \int \theta(V-V_2) p(V) dV  \right] }^{N-2} \, .
	\end{eqnarray}
Here the integration goes over the sector $V_1 < V_2$ while the contribution of the
sector $V_2 < V_1$ is accounted by the factor 2 before the integral. Twice applying
the relation
\begin{eqnarray}
		\label{relation}
p(V)\left[\int \theta(V'-V)p(V') dV'\right]^M = -\frac{1}{M+1} \nonumber \\ 
\times \frac{d}{dV}
\left[\int \theta(V'-V)p(V') dV'\right]^{M+1}
	\end{eqnarray}
we arrive at Eq.(\ref{pb0}). Naturally, this combinatoric result holds for an arbitrary
distribution function $p(V)$. To find the linear in $U$ correction $P_{b}^{(\mathrm{lin})}$
to Eq.(\ref{pb0}) one
should sequentially expand the theta-functions in the integrand of Eq.(\ref{p12general}),
$\theta(V+U) \rightarrow \theta(V) + U\delta(V)$.
Summing up the contributions from all theta-functions, we obtain $P_{b}^{(\mathrm{lin})}(U)$
in the following form:
	\begin{eqnarray}\label{p(1)}
		P_{b}^{(\mathrm{lin})}(U) &=&  2N(N - 3)U
		\int dV_1 p(V_1) \int dV_2 \theta(V_2-V_1)  \nonumber \\
& &  \times p^2(V_2) \left[ \int dV \theta(V_2 -V) \right]^{N-3}.
	\end{eqnarray}
Because of the delta-function, the probability distribution for one of the sites
is squared, and, quite expectably, the integral depends on the explicit
shape of the distribution. For the considered box distribution  Eq.(\ref{eq:p_V})
we obtain:
\begin{equation}\label{p(1)-answer}
		P_{b}^{(\mathrm{lin})}(U) = \frac{2N(N-3)}{(N-2)(N-1)}U \, .
	\end{equation}
The coefficient by $U$ tends to 2 for large $N$.

\begin{figure}[h]
		\centering
		\includegraphics[width=0.9\linewidth]{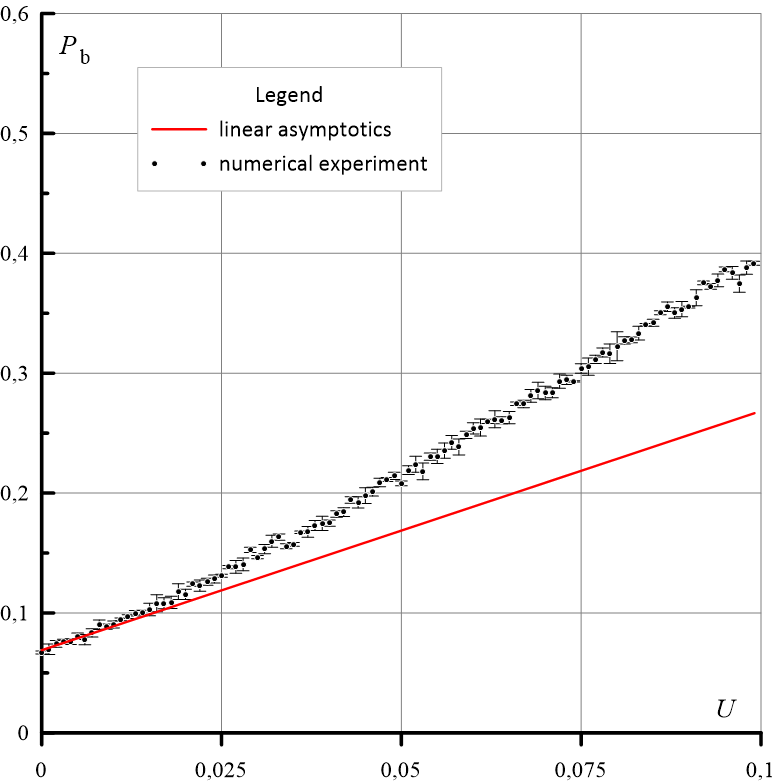}
		\caption{Dependence of the bound state probability $P_\text{b} (U)$ on the relative interaction strength $U$ in the range of $U \ll 1$ for the number of sites  $N =30$.}
		\label{fig:linear-asympts}
	\end{figure}

 	Putting the obtained linear dependences (\ref{p(1)-answer}) in the results of the numerical experiment (Fig. \ref{fig:linear-asympts}), we observe that the linear asymptotics are true in the region $U \ll 1/N$ but deviate from the experiment at $1/N \leq U$.

	\subsection{Crossover}\label{crossover}
A wider range of the interaction strength $1/N \ll U \ll 1$ can be described with the following
reasoning. The average energy distance between $N$ on-site potentials, randomly distributed
over the unit energy interval, is $1/N$. An average number of sites with random potentials
$V < \eta U$, where $\eta \lesssim 1$, is given by $K = \eta U N$ and obeys the inequality
$1\ll K \ll N$. If two of such sites are neighboring, the fermion pair located on them will
certainly (for $\eta < 1/2$ ) or with a good probability (if $1/2 < \eta \lesssim 1$) have a
negative energy, i.e., it will be bound.
The probability $P_b$ that a fermion pair is bound can be represented as $P_b = 1 - P_{s}$ where
$P_{s}$ is the probability that the two fermions are separated, that is among the $K$ sites
there are no neighboring ones.
This elementary combinatorial problem gives (in the limit $1\ll K \ll N)$: $P_{s} = \exp(-K^2/N)$.
Combining with Eqs.(\ref{pb0}) and (\ref{p(1)-answer}), we arrive at the interpolation formula
for $P_b(U)$ at $U \ll 1$:
% in the long chain ($N \gg 1$):
\begin{eqnarray}
		P_\text{b}(U) \approx \frac{2}{N} + 2U + 1 - e^{- \eta^2NU^2} \approx  \eta^2 N U^2 \, , \label{P-intermediate}
\end{eqnarray}
where the last equality corresponds to the narrower intermediate interval $1/N \ll U \ll 1\sqrt{N}$
and shows the dominant quadratic dependence of the function $P_b(U)$ on $U$ presented in Fig. \ref{fig:linear-asympts}. Although the
numerical constant $\eta \lesssim 1$ still remains  uncertain (it will be fixed in section \ref{TM}), Eq.(\ref{P-intermediate})
allows to infer that the crossover between the regimes of almost decoupled and almost coupled
fermions occurs at $U \sim 1/\sqrt{N}$. These behaviors agree with the numerical experiment (dots in Fig. \ref{fig:crossover}).

Note in passing that the above qualitative analysis of the crossover 
can be extended to a cluster of an arbitrary dimension $d$ resulting
in the expression $P_s = \exp\left(-d\eta^2U^2N\right)$. This allows us to 
conjecture that the estimate $U \sim 1/\sqrt{N}$ for the crossover range 
is universal. A regular description of the crossover will be developed below 
for the one-dimensional cluster (chain) by the transfer matrix approach.

 	\section{Transfer matrix method}\label{TM}
 \subsection{Basic equations}\label{basic_eqs}
Consider the case where the ground state corresponds to separated fermions, i.e.,
occupying two non-neighboring sites, e.g., the $1^{st}$ and the $k^{th}$ ones
($k \neq 2, N$). Due to equivalence of configurations $(1, k)$, $(2, k+1)$, etc., the probability $P_s(1,k)$
of this particular arrangement is simply connected with the total probability
$P_s = 1-P_b$ of the realization of separated fermions:
\begin{equation}\label{Ps-sum}
P_s = \frac{N}{2}\sum^{N-1}_{k=3}P_s(1,k).
\end{equation}
To correspond to the ground state, the energy of the configuration with fermions
located on the $1^{st}$ and the $k^{th}$ sites, $V_1 + V_k$, should obey to a set of
obvious inequalities, like $V_1 + V_k < V_i + V_{i+1}-U$,
$V_1 + V_k < V_i +V_k$ ($i \neq 1, k -1, k, k+1)$, etc. The probability $P_s(1,k)$
is determined by the averaging of these inequalities over the disorder:
\begin{widetext}
		\begin{eqnarray}
			\label{eq:Ps_general}
			P_\text{s}(1,k)(U) &=& \bigg\langle
\theta \left(V_N - \varepsilon_1 \right)
\prod_{i=k+1}^{N-1} \left[ \theta \left(V_{i+1} - \varepsilon \right)
\theta \left(V_{i+1} + V_{i} - {U} - E  \right)\theta \left(V_{i} - \varepsilon \right) \right]
\, \theta \left(V_{k+1} - \varepsilon_k \right)
\nonumber  \\
& \times & \, \theta \left(V_{k-1} - \varepsilon_k\right)
\prod_{i=2}^{k-2} \left[ \theta \left(V_{i+1} - \varepsilon \right) \theta \left(V_{i+1} + V_{i} - {U} - E  \right)\theta \left(V_{i} - \varepsilon \right) \right]
\,  \theta \left(V_2 - \varepsilon_1 \right)
			   \bigg\rangle.
		\end{eqnarray}
	\end{widetext}
Here we have introduced several parameters $\varepsilon_1$, $\varepsilon_k$,  $E$ and $\varepsilon$,
determined by the potentials $V_1$ and $V_k$:
\begin{eqnarray}
		&	\varepsilon_1 = \max \left\{ V_1, V_k + {U} \right\} , \,\, \varepsilon_k = \max \left\{ V_k, V_1 + {U} \right\}, \nonumber \\
		&	E = V_1 + V_k  ,  \,\, \varepsilon = \max \left\{ V_1, V_k \right\}.
			\label{eq:E_eps}
\end{eqnarray}
 The expression (\ref{eq:Ps_general}) is represented in the form suitable for the celebrated
 transfer matrix approach \cite{Kramers}. For this aim we introduce the operator $\hat{A}$
	\begin{eqnarray}
		\label{eq:A-gen}
		\hat{A} \left(V, V' \right) &=& \sqrt{p(V)}\theta \left(V - \varepsilon \right) \theta \left(V + V' - {U} - E  \right) \nonumber \\
& \times &  \theta \left(V' - \varepsilon \right)\sqrt{p(V')} \,
	\end{eqnarray}
and note that the matrix product
	\begin{equation}
		\label{eq:A-gen-prod}
		\hat{A}^2\left(V, V' \right) =
\int \hat{A} \left(V, V^{''} \right)\hat{A} \left(V^{''}, V' \right) dV^{''}
	\end{equation}
just corresponds to the product of two subsequent blocks in Eq.(\ref{eq:Ps_general}) with
averaging over the intermediate variable, i.e., with the integration with the weight
$p(V^{''})$. Thus, $\hat{A} \left(V_{i+1}, V_{i} \right)$
is the desired transfer matrix between the sites $i$ and ${ i+1 }$ for the chain
with an arbitrary disorder distribution function $p(V)$. Postponing the study of this general
case for future, in the present article we concentrate on the
particular case of the box distribution (\ref{eq:p_V}), so
\begin{eqnarray}
\label{eq:A}
\hspace{-0.2cm} \hat{A} \left(V, V' \right) = \theta \left(V - \varepsilon \right) \theta \left(V + V' - U - E  \right) \theta \left(V' - \varepsilon \right) .
	\end{eqnarray}
Note that $\hat{A}$ depends on the parameters $E$ and $\varepsilon$, determined entirely by the potentials on the two selected sites, see Eq.(\ref{eq:E_eps}).

Using Eqs.(\ref{eq:A-gen-prod}) and (\ref{eq:A}), the probability (\ref{eq:Ps_general}) can be rewritten in the form
\begin{eqnarray}
	\label{eq:Ps_transfer}
	&&\hspace{-0.5cm} P_\text{s}(1,k)=\bigg\langle \theta \left(V_N - \varepsilon_1 \right) \hat{A}^{N-k-1} \left( V_N, V_{k+1}  \right) \theta \left(V_{k+1} - \varepsilon_k \right)  \nonumber
	\\ 
	 &&\times \, \theta\left(V_{k-1} - \varepsilon_k\right)\hat{A}^{k-3} \left(V_{k-1}, V_2 \right) \theta \left(V_2 - \varepsilon_1 \right) \bigg\rangle .
\end{eqnarray}
Here the first and the second lines have resulted from the integration
over $V_{N-1}, \ldots, V_{k+2}$ and $V_{k-2}, \ldots, V_{3}$, respectively. The remaining averaging over the disorder is reduced to the integration over the potentials on the $1^{st}$ and the $k^{th}$ sites, and on their nearest neighbors.

Being real and symmetric in $V$ and $V'$, the transfer matrix $\hat{A}$ can be represented as
\begin{equation}\label{A-repr}
		\hat{A}\left(V, V' \right) = \sum_{\nu} \varphi_{\nu} \left(V\right) \lambda_{\nu}
\varphi_{\nu} \left(V'\right) \, ,
	\end{equation}
where the eigenfunctions $\varphi_{\nu} \left(V\right)$ obey the equation
$\hat{A}\varphi_{\nu} = \lambda_{\nu} \varphi_{\nu}$ and constitute an orthonormal basis.
Like the matrix $\hat{A}$, the eigenfunctions and the eigenvalues depend on the `external' parameters $E$
and $\varepsilon$. Powers of $\hat{A}$ are given by
\begin{equation}\label{eq:A^M_general}
\hat{A}^M\left(V, V' \right) = \sum_{\nu} \varphi_{\nu} \left(V\right) \lambda^M_{\nu}
\varphi_{\nu} \left(V'\right) \, .
	\end{equation}
For large $M$ the leading term in the sum (\ref{eq:A^M_general}) is that with
the largest modulus eigenvalue, say, $\lambda_0$. The central point of the transfer matrix
method is to replace $\hat{A}^M$ by its leading part:
	\begin{equation}\label{A-appr}
   \hat{A}^{M}\left(V, V' \right) \rightarrow \varphi_0 \left(V\right) \lambda_0^{M}
    \varphi_0 \left(V'\right).
	\end{equation}
In our case of a long chain, the leading contribution to the sum (\ref{Ps-sum})
is provided by sites with $k \sim N \gg 1$, so the transfer-matrix method is appropriate.
Applying Eq.(\ref{A-appr}) to Eq.(\ref{eq:Ps_transfer}) we see
that in this limit the contribution of a particular pair of sites (like $(1,k)$) does not
depend on $k$. Therefore, in the leading order in $N$ we obtain
\begin{eqnarray}
			\label{eq:Ps_transfer-simple}
			P_\text{s} &=& \frac{N^2}{2}
\int^1_0\int^1_0 dV_1 dV_k \lambda_0^{N-4} I^2(\varepsilon_1)I^2(\varepsilon_k) \, ,
		\end{eqnarray}
where the function $I(e)$ is defined by
\begin{eqnarray}\label{I-e}
I(e) = \theta(1-e)\int\limits_{e}^{1} \varphi_0 \left( V \right) dV \, .
\end{eqnarray}
The integrand in Eq.(\ref{eq:Ps_transfer-simple}) is an implicit function
of the variables $V_1$ and $V_k$ [see  Eq.(\ref{eq:E_eps})], subject to
the constraints imposed by Eq.(\ref{I-e}):
\begin{equation}
	\label{eq:external_con}
	\varepsilon_1 \, , \, \varepsilon_k <1 \,\, \Rightarrow \,\, V_1 \, , \, V_k < 1 - U.
\end{equation}
To proceed we need to find the largest eigenvalue $\lambda_0$ and the
corresponding eigenfunction $\varphi_0 \left( V \right)$.

\subsection{Solution of the integral equation $\hat{A}\varphi_{\nu} = \lambda_{\nu} \varphi_{\nu}$}\label{solution}

In accordance with Eq.(\ref{eq:A}),  the integral equation
 \begin{equation}
		\label{eq:int_eq}
		\int\limits_{0}^{1} \hat{A} \left(V, V' \right)  \varphi \left(V'\right) d V' = \lambda  \varphi \left(  {V}\right).
	\end{equation}
is actually restricted to the region $\varepsilon < V, V' < 1$.  Eq. (\ref{eq:int_eq})
has different forms in the three areas of possible relations between the `external' parameters
$E$ and $\varepsilon$ (i.e., between $V_1$ and $V_k$). We write down these three equations in
the allowed region $\varepsilon < V, V' < 1$:\\
\textbf{1.} $\Sigma_1$: $E + {U} - \varepsilon < \varepsilon < 1 $ (here the argument of the middle theta function in
Eq.(\ref{eq:A}) is positive):
\begin{equation}
		\label{eq:region1}
		\lambda  \varphi \left(  {V}\right) = \int\limits_{\varepsilon}^{1}   \varphi \left(V'\right) d V' \, ;
	\end{equation}
\textbf{2.} $\Sigma_2$: $\varepsilon < E + {U} - \varepsilon < 1 $:
	\begin{eqnarray}
		\label{eq:region2}
		\lambda  \varphi \left(  {V}\right)&=& \theta \left(E + {U} - \varepsilon -  V\right)
        \int\limits_{E + {U} - V}^{1}   \varphi \left(V'\right) d V'
		\nonumber \\
		&+& \theta \left( V-E - {U} + \varepsilon \right)
        \int\limits_{\varepsilon}^{1}   \varphi \left(V'\right) d V' \, ;
	\end{eqnarray}
\textbf{3.} $\Sigma_3$: $\varepsilon < E + {U} - 1 < 1$:
	\begin{equation}
		\label{eq:region3}
		\lambda  \varphi \left(  V\right) =
\theta \left( V-E - {U} + 1 \right)  \int\limits_{E + {U} - V}^{1}   \varphi \left(V'\right) d V'.
	\end{equation}
Note at once that the solution in the third area does not contribute to Eq.(\ref{eq:Ps_transfer-simple})
due to vanishing of the functions $I(\varepsilon_1)$ and $I(\varepsilon_2)$, Eq.(\ref{I-e}). Indeed, as follows from the left inequality in the definition of the third area, $1-U < E-\varepsilon = \min\{V_1, V_k\}$. This contradicts  the `external' constraint (\ref{eq:external_con}). Thus, we need to consider only the two remaining areas.

\begin{figure}[] 
	\centering
	\includegraphics[width=1\linewidth]{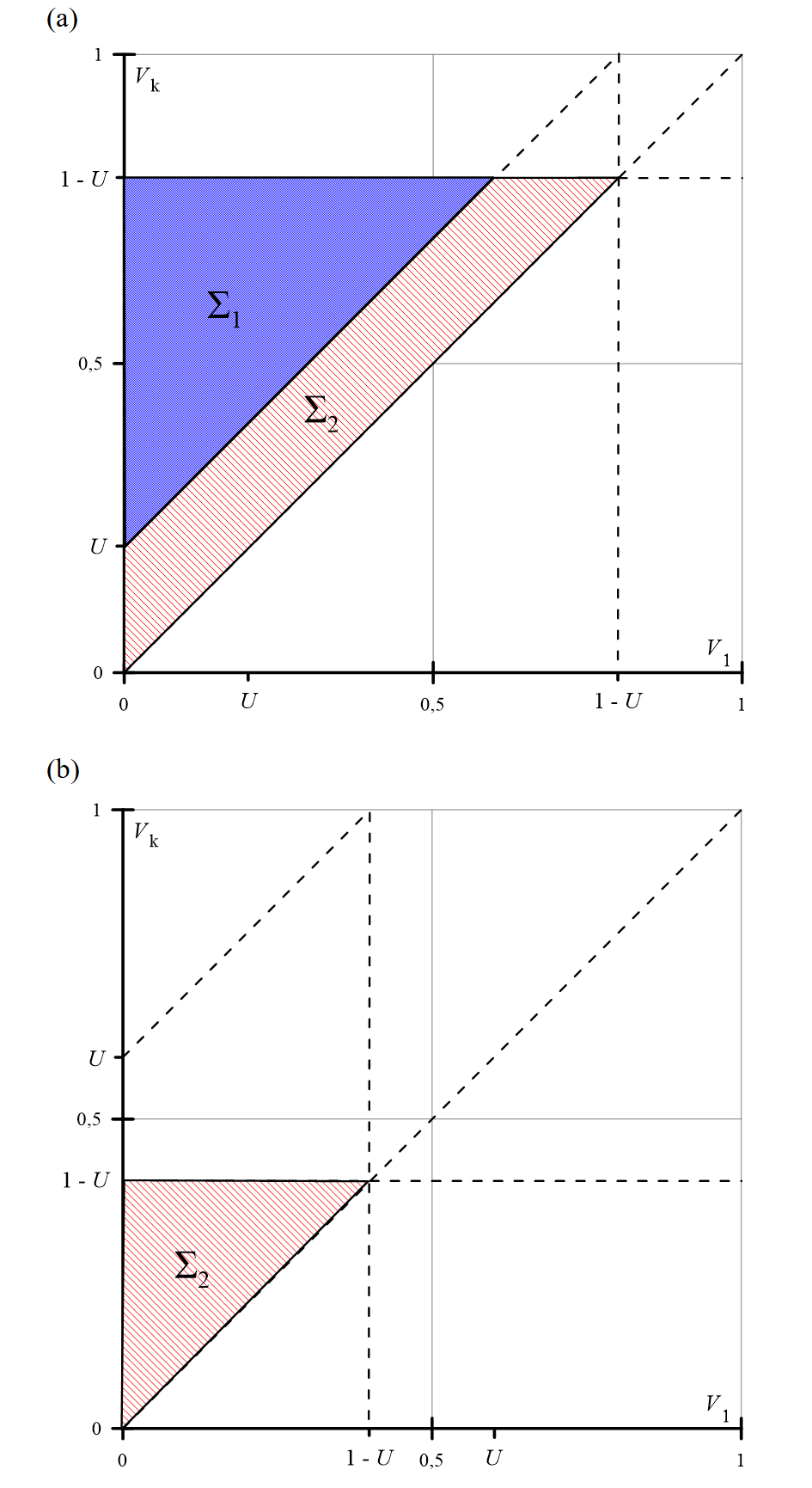}
	\caption{The areas $\Sigma_1$ and $\Sigma_2$, determined by the conditions (\ref{eq:region1}) and (\ref{eq:region2}), on the plane of the `external' variables $V_1$ and $V_k$ (in the sector $V_1 < V_k$) for the interaction strength $U < 1/2$ (a) and $U \geq 1/2$ (b).}
	\label{fig:areas}
\end{figure}

\begin{figure}[H] 
	\centering
	\includegraphics[width=0.8\linewidth]{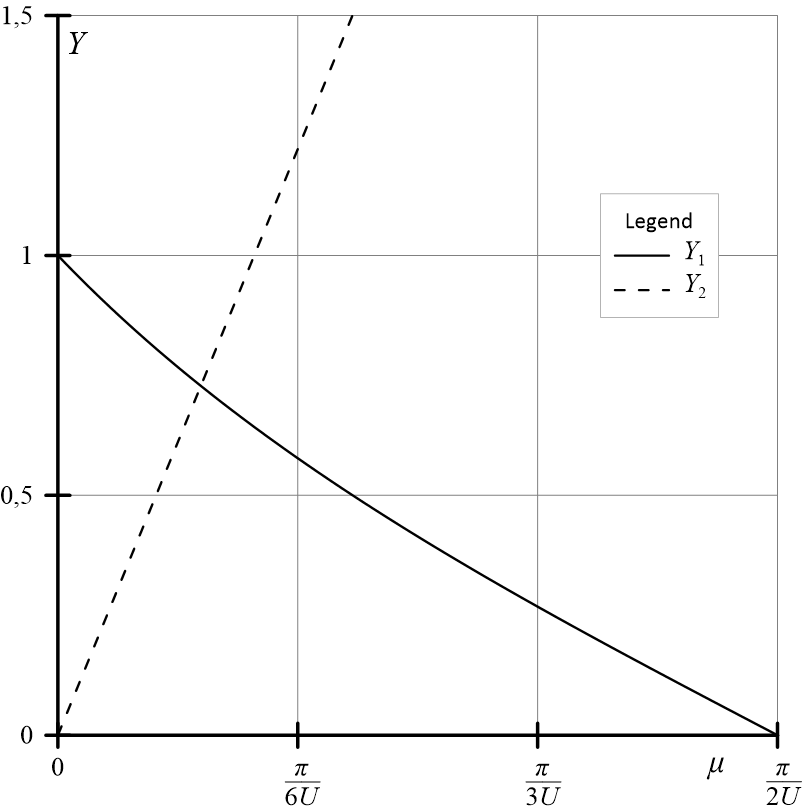}
	\caption{Graphical representation of the characteristic equation (\ref{eq:tan}). It corresponds to the intersection of the curve $Y_1 = \tan[ \pi/4 - \mu \left(U+V_1 - V_k \right)/2]$
		and the straight line  $Y_2 = \mu (1 - U - V_1)$. }
	\label{fig:Y}
\end{figure}

In the first area, $\Sigma_1$, the equation (\ref{eq:region1}) possesses a single solution
(the superscript marks the area):
\begin{eqnarray}
	\label{eq:ph1}
	\varphi^{(1)}(V) &=& \dfrac{1}{\sqrt{1-\varepsilon}}\theta \left(V - \varepsilon \right)
\theta \left(1 - V \right) \\
\lambda^{(1)} &=& 1 - \varepsilon ,
\end{eqnarray}
which means that the matrix $A$ in this area reduces to a projector. However, the area
$\Sigma_1$ contributes to the integral Eq.(\ref{eq:Ps_transfer-simple}) only when $U < 1/2$. It becomes obvious if to rewrite the left inequality in the condition defining this area
in the form $E+U-\varepsilon = \min\{V_1,V_k\} +U < \varepsilon = \max\{V_1,V_k\}$ and
to account for the `external' requirement $\max\{V_1,V_k\} < 1 - U$ (\ref{eq:external_con}).
For the case $U < 1/2$, the sector $V_1 < V_k$ of $\Sigma_1$ is depicted in
Fig. \ref{fig:areas}{(a)}. Integrating in (\ref{eq:Ps_transfer-simple}) over $\Sigma_1$
we find the contribution $P_\text{s}^{(1)}$ of this area to the probability $P_\text{s}$:
\begin{eqnarray}
		\label{eq:Ps_region1}
		P_{\text{s}}^{(1)}= \left(1-{U}\right)^{N} {\left( \frac{1 - 2{U}}{1 - U} \right)}^2. % \sim e^{-UN}.
	\end{eqnarray}
In the second area $\Sigma_2$, the eigenfunctions of the integral equation (\ref{eq:region2}) have the following form
	\begin{eqnarray}
	\label{eq:area2_solution_general}
\varphi^{(2)}(V) = \left[ A \cos\left(\mu V\right) \right. & + & \left. B \sin\left(\mu V\right) \right] \theta\left( E + {U} - \varepsilon - V\right) \nonumber \\
& + & C \,\theta\left(V - E - {U} + \varepsilon\right) \, .
	\end{eqnarray}
Here, for convenience, the notation $\mu$ has been introduced
for the inverse eigenvalue
\begin{equation}
	\mu = 1 / \lambda,
\end{equation}
so we need to find solutions with the smallest modulus of $\mu$.
Eq.(\ref{eq:region2}) imposes the set of conditions on the coefficients $A, B, C$:
\begin{eqnarray} \label{eq:ABC}	
	C \mu \left( 1 - E -  {U}  + \varepsilon \right) &=&  B \cos\left( \left( E + {U} - \varepsilon\right) \mu \right) \nonumber
	\\
	& &- A \sin\left( \left( E + {U} - \varepsilon\right) \mu \right), \nonumber
	\\
	A\left[1 + \sin\left(\left(E + {U} \right) \mu \right) \right] &=& B \cos \left(\left(E + {U} \right) \mu \right), \nonumber
	\\
	B \left[1 - \sin\left(\left(E + {U} \right) \mu \right) \right] &=& A \cos \left(\left(E + {U} \right) \mu \right), \nonumber
	\\
	B \cos\left( \varepsilon \mu \right) - A \sin\left( \varepsilon \mu \right) &=& C,
\end{eqnarray}
where the second and the third linear equations are equivalent
(the determinant of this subsystem is identically zero).
The system (\ref{eq:ABC}) leads to the characteristic equation for $\mu$
\begin{equation}
	\label{eq:tan}
	\left( 1 - E -  {U}  + \varepsilon \right) \mu = \tan \left(\frac{\pi}{4} - \frac{\left( E + {U}-2\varepsilon\right) \mu }{2}  \right) \, ,
\end{equation}
that has an infinite number of solutions for a given set of parameters $E$, $\varepsilon$, and $U$.
The solution of our interest $\mu_0$ with the minimal modulus lies in the interval, where the tangent argument
varies between $0$ and $\pi/4$, see Fig. \ref{fig:Y}. 

The coefficients $A$ and $B$ can be expressed via $\mu_0$ and $C$
with $C$ determined by the normalization condition 
\begin{equation}\label{eq:C2-norm}
	\int\limits_{0}^{1}   {\left|\varphi^{(2)}_0\left(V\right)\right|}^2 dV = 1 \, .
\end{equation}
The coefficient $C$ also determines the functions $I(\varepsilon_1)$ and $I(\varepsilon_k)$ in Eq.(\ref{eq:Ps_transfer-simple}). Indeed, both $\varepsilon_1$ and $\varepsilon_k$ are greater
than $E+U - \varepsilon$, hence the eigenfunction $\varphi^{(2)}(V)$ in the integrand of Eq.(\ref{I-e})
is just $C$ [see Eq(\ref{eq:area2_solution_general})], so
\begin{eqnarray}\label{I-e-2}
	I(\varepsilon_1) = \left( 1 - \varepsilon_1 \right) C\,\,  ,  \,\,
I(\varepsilon_k) = \left( 1 - \varepsilon_k \right) C
\, .
\end{eqnarray}
Both $C$ and $\mu_0=1/\lambda^{(2)}_0$ are functions of $U$ and variables $V_1, V_k$
within the area $\Sigma_2$. This area exists for any $U$ (from the interval $0 < U < 1$)
and is depicted (for the sector $V_1 < V_k$) in Fig. \ref{fig:areas}{(a)}
and Fig. \ref{fig:areas}{(b)} for $U< 1/2$ and $1/2 < U$, respectively.

The transcendental equation (\ref{eq:tan}) for $\mu$ can be solved only numerically.
In general, this makes  a straightforward analytical calculation of the integral
(\ref{eq:Ps_transfer-simple}) over the area $\Sigma_2$ impossible.
However, an analytical calculation is possible in the limit of our interest $N \gg 1$. The leading
in $N$ contribution to the integral over $\Sigma_2$ is given by a vicinity of the point
$(V_{1*}, V_{k*})$ where $\lambda^{(2)}_0(V_1, V_k)$ has a maximum, i.e., $\mu_0(V_1, V_k)$
is minimal (in modulus).

\begin{figure}[h] 
	\centering
	\includegraphics[width=0.8\linewidth]{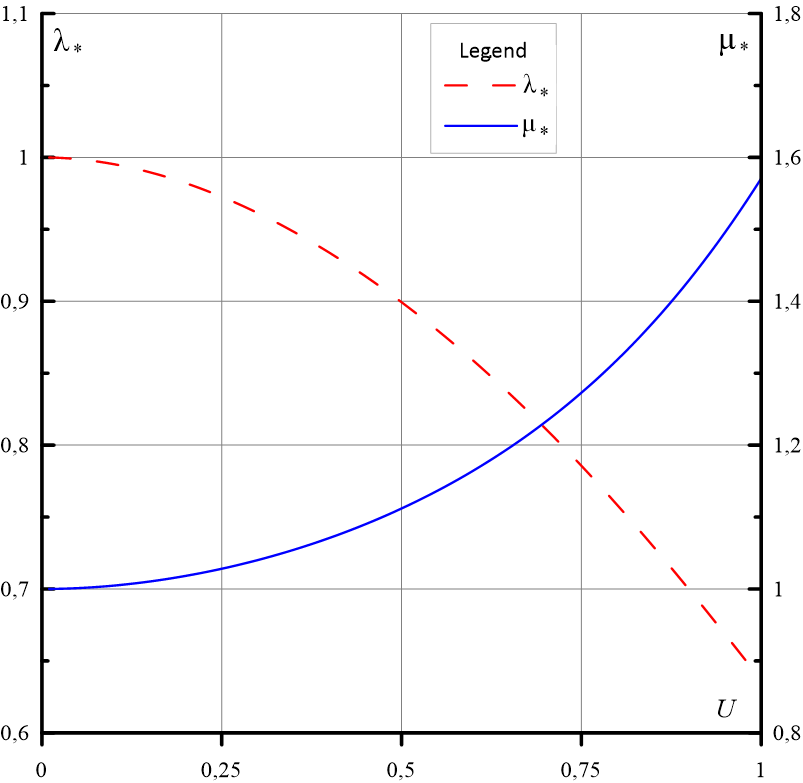}
	\caption{Dependence of $\mu_*$ and $\lambda_* = 1/\mu_*$ on $U$ 
		given by the numerical solution of Eq. (\ref{eq:tan_0}).}
	\label{fig:mu}
\end{figure}

This minimum $\mu_{*}$ is reached at the point $(V_1=0, V_k=0)$.
To prove this statement consider, for certainty, the sector $V_1 < V_k$ and note that
the value of $\mu$ is determined graphically on the plane
($\mu$, $Y$) by the intersection of the curve $Y_1 = \tan[ \pi/4 - \mu \left(U+V_1 - V_k \right)/2]$ and the straight line $Y_2 = \mu (1 - U - V_1)$ (Fig. \ref{fig:Y}). With a fixed slope of the straight line $Y_2$ (i.e., fixed $V_1$ ), the intersection of the curve and the straight line occurs at a smaller coordinate $\mu$, the closer the right end of the interval is to the origin. So, the minimum happens at the minimal $V_k$ that corresponds to the case $V_k = V_1$. Further, note that  the steeper the line $Y_2$,  the smaller the intersection coordinate $\mu$  is. The steepness increases with the decrease of $V_k$ and becomes maximal at $V_k = 0$.
This proves the announced statement that the minimal value $\mu_*$ is determined by the numerical solution of the equation
\begin{equation}
	\label{eq:tan_0}
	\left( 1  -  {U}  \right) \mu = \tan \left(\frac{\pi}{4} - \frac{  U \mu }{2}  \right).
\end{equation}
Its solution $\mu_* (U)$ together with $\lambda_*(U) = 1/\mu_* (U)$  are plotted as the functions of $U$ in Fig. \ref{fig:mu}.
In a close vicinity of the point $(V_1=0, V_k=0)$ the function $\mu_0(V_1, V_2)$ can be
represented as $\mu_0(V_1, V_k)= \mu_{*} + \delta \mu$, where the leading order correction,
determined by Eq.(\ref{eq:tan}), is given by
	\begin{eqnarray}
		&& \delta \mu(V_1, V_k) = \gamma_1 V_1 + \gamma_k V_k  \, ,  \label{eq:delta_mu} \\
		\gamma_{1(k)} &=& \frac{1 \mp (1 - U)^2\mu^2_{*}}
		{2(1 - U)  + U \left[ 1 + \left( 1 - U \right)^2\mu^2_{*}\right] } \, ;  \label{gamma}
	\end{eqnarray}
here the upper (lower) sign relates to $\gamma_1$  ($\gamma_k$).
Correspondingly, the $N^{\text{th}}$ power of the maximal eigenvalue $\lambda^{(2)}_0 = 1/(\mu_* + \delta \mu)$ in
a vicinity of the point $(V_1=0, V_k=0)$ can be represented as 
\begin{equation}
	\label{eq:lambda}
	[\lambda^{(2)}_0(V_1, V_k)]^N =
\lambda^N_* \exp\left( - \frac{N\delta \mu(V_1, V_k)}{\mu_*}\right) \, ,
\end{equation}
where $\lambda_* = \lambda^{(2)}_0(0, 0) = 1/\mu_{*}$. As is seen from Eq.(\ref{eq:lambda}) typical values of $V_1, V_k \sim 1/N \ll 1$, which justifies the linearization of the characteristic equation.
This smallness allows us to take the functions $I(\varepsilon_1), I(\varepsilon_k)$
in the integrand of Eq.(\ref{eq:Ps_transfer-simple}) at $V_1,V_2 =0$. Using Eq.(\ref{I-e-2}), we
get $I(\varepsilon_1)= I(\varepsilon_k)=C$, where the normalization constant $C$, determined by
Eq.(\ref{eq:C2-norm}) at $V_1,V_2 =0$,  equals
\begin{equation}
	\label{eq:C2}
	C^2 = \frac{2}{2 - U + U {\left( 1 - U \right)}^2 \mu_*^2}.
\end{equation}
Performing the integration in Eq.(\ref{eq:Ps_transfer-simple}) we arrive at the final expression
for the contribution $P_{\text{s}}^{(2)}$ from the area $\Sigma_2$ to the
probability $P_{\text{s}}$:
\begin{equation}
		\label{eq:area2_result}
		P_{\text{s}}^{(2)}(U) = \frac {2 \left( 1 - U \right)^4 \lambda_*^{N-2}(U)}
		{\lambda_*^2(U) + \left(1-U\right)^2 }\left[1 - e^{-\gamma_k NU}  \right] \, .
\end{equation}
At $1/N << U$, the term $\exp\left(-\gamma_k NU\right)$ is negligible and can be omitted.
	We deliberately keep this term for the further discussion of the range of ultra-small $U \ll 1/N$.

Eq.(\ref{eq:area2_result}) holds for almost all values of $U$ excepting
	the narrow region $1-U \ll 1/N$, when the domain $\Sigma_2$ shrinks to a tiny triangle,
	see Fig. \ref{fig:areas}{(b)}. In the integral (\ref{eq:Ps_transfer-simple}) over this triangle we may
	neglect the dependence of the eigenstate $\lambda_0$ on the coordinates, but we have to
	account for this dependence in the functions (\ref{I-e}):
	$I(\varepsilon_1)=C(1-U-V_k)$ and $I(\varepsilon_k)=C(1-U-V_1)$
	[see Eq.(\ref{I-e-2}) in the sector $V_1 < V_k $]. As a result, we obtain:
\begin{equation}
		\label{eq:area2_result-U1}
		P_{\text{s}}^{(2)}(U) = \frac {2N^2}{9}(1-U)^6 \lambda_*^{N-4}(U) \, \, \, ; \,\, \, 1-U \ll \frac{1}{N} \, ,
 \end{equation}
	where $\lambda_*(U = 1) = 2/\pi$.

\subsection{Results and discussion}\label{results}
The expressions (\ref{eq:Ps_region1}), (\ref{eq:area2_result}), and (\ref{eq:area2_result-U1}) completely determine
	the probability $P_b(U) = 1 - P_{\text{s}}^{(1)}(U)	- P_{\text{s}}^{(2)}(U)$
	for two fermions to form a `bound pair' (i.e., to be localized on neighboring sites)
	in the ground state of a long strongly disordered chain. Except in the narrow
	region $1-U \ll 1/N$, $P_b(U)$ is given by:
\begin{eqnarray}
	\label{eq:Pb_final result}
	P_b(U) &=& 1 - \theta(1/2 - U)\left(1-{U}\right)^{N} \left( \frac{1 - 2{U}}{1 - U} \right)^2
	\nonumber \\
	&-& \frac{2 \left( 1 - U \right)^4 \lambda_*^{N-2}(U)}
	{\lambda_*^2(U) 
		+ \left(1-U\right)^2 }\left[1 - e^{-\gamma_k NU}  \right] \, .
\end{eqnarray}

The contributions from the areas $\Sigma_1$ and $\Sigma_2$ are mostly determined 
by the high powers of $1-U$ and $\lambda_*$, respectively.
According to the characteristic equation (\ref{eq:tan_0}), the minimal 
inverse eigenvalue $\mu_*$ obeys the inequality $(1-U)\mu_* < 1$
(because the tangent function is less than unity at the 
interval of interest, see Fig. \ref{fig:Y}). This means $1-U < \lambda_*$ 
and at not too small $U$ the contribution $P_{\text{s}}^{(1)}(U)$ is much smaller than $P_{\text{s}}^{(2)}(U)$; the latter is also small and decreases exponentially with the increase of $N$. However, the both eigenvalues, $1-U$ and $\lambda_{*}(U) \approx 1- U^2/2$ tend to unity when $U \ll 1$. As we shall see, the crossover from bound to decoupled fermions occurs just at small $U \ll 1$, so this range deserves a particular attention.

As shown in section \ref{weak_interaction} , in the regime of ultra-low
$U \ll 1/N$, the interaction gives
only a tiny correction (\ref{p(1)-answer}) to the trivial
combinatoric expression (\ref{pb0}). 
The both terms are written in the limit of large $N$ and
are beyond the accuracy of our subsequent analysis. To illustrate
the consistency of the latter, note that `large' linear terms ($\sim NU$)
of the formal expansion of the two terms in Eq.(\ref{eq:Pb_final result})
mutually cancel. The fermions may be considered as almost decoupled in this
regime.

At larger $U$, when $1/N \ll U \ll 1/N^{1/4}$, the term $P^{(1)}_s \propto \exp\left[N\ln\left(1-U\right)\right]$ becomes exponentially small,
while the term $P^{(2)}_s \approx \exp\left(-NU^2/2\right)$ varies between
unity and (almost) zero. The binding probability $P_b(U)$ in this range
\begin{eqnarray}
	\label{eq:Pb_final result-cross}
	P_b(U)= 1 - e^{-NU^2/2} \, .
\end{eqnarray}
describes the crossover at
\begin{eqnarray}
	\label{eq:crossover}
	U \sim 1/\sqrt{N} \,
\end{eqnarray}
from the regime of almost decoupled fermions
(at $1/N \ll U \ll 1/N^{1/2}$) to the almost bound ones
(at $1/N^{1/2} \ll U$. Note that the calculated functional dependence
of $P_b(U)$ in this regime coincides with that obtained
with the qualitative reasoning in section \ref{crossover} [see the paragraph above
Eq.(\ref{P-intermediate})] and fixes the value of the
unknown constant: $\eta^2 = 1/2$.

At still larger values of $U$ both $P^{(1)}_{\text{s}}(U)$ and $P^{(2)}_{\text{s}}(U)$
decay exponentially with the increase of $N$, so the binding probability $P_b$ approaches
unity, while the decoupling of fermions in a long chain becomes a rare event.
At $U<1/2$, the decoupling probability $P_{\text{s}} = P^{(1)}_{\text{s}} + P^{(2)}_{\text{s}}(U)$
is determined by the contributions from the both areas $\Sigma_1$, (\ref{eq:region1}), and $\Sigma_2$, (\ref{eq:region2}); the latter contribution dominates.
At $1/2 < U$, the decoupling probability is given entirely by Eq. (\ref{eq:area2_result}) and
is an exponentially decaying function of $N$: 
$P^{(2)}_{\text{s}}(U) \propto \exp\left[-N\ln(1/\lambda_{*}(U))\right]$ with $\lambda_{*}(U)$ 
changing from $\approx 0.9$
to $2/\pi$ when $U$ changes from $1/2$ to $1$. At a fixed number of sites, $P_{\text{s}}(U)
	\propto (1-U)^4 \rightarrow 0$ when $U$ approaches unity but still lies outside the narrow region
	$1-U \ll 1/N$, where this dependence changes into $P_{\text{s}}(U) \propto N^2(1-U)^6$.

All these analytical results are confirmed by numerical experiments, see Fig. \ref{fig:crossover}.

\section{Conclusion and open problems}\label{conclusion}

We have described a disorder-induced decoupling of a pair of identical fermions with a short-range attractive interaction on a finite lattice cluster with random on-site energies. In contrast to attracting nonidentical fermions (e.g., with different spins), which can simultaneously occupy a site with a minimal energy and thus always form a bound state resistant to disorder, for the identical fermions the probability $P_b$ of pairing on neighboring sites depends on the relation between the interaction $\widetilde{U}$ and the disorder distribution width $W$ (both $\widetilde{U}$ and $W$ are assumed to be large as compared to the
kinetic energy, i.e., the intersite hopping rate, so the system is deeply in the regime 
of the single-particle Anderson localization \cite{Anderson1958}).

For a cluster of  arbitrary dimension, we have presented a qualitative argument for a crossover between the regimes of almost coupled and almost decoupled configurations in 
the ground state. This crossover takes place at $\widetilde{U}/W \sim 1/\sqrt{N}$,
where $N$ is the number of lattice sites ($N \gg 1$). 
However, a straightforward brute-force analytical calculation or computation of the pairing 
probability $P_b$ as a function of $\widetilde{U}$ and $W$ is an arduous task even for the 
simplest cluster in the form of a closed chain and for the simplest box-like 
distribution of the disorder. The latter problem turns out to be equivalent to the 
computation of the volume of a polyhedron (in general, NP-hard). 

Remarkably, we have found that in the chain geometry the problem can be solved by the transfer matrix method. In the case of the box-distribution of the disorder, the eigenvectors and the eigenvalues of the transfer matrix can be derived analytically (another wonder!) from a rather non-trivial integral equation. Using this approach we have calculated the pairing probability in the long chain for an arbitrary relation between the interaction and the disorder strengths. In particular, we have explicitly described the coupling-decoupling crossover. The obtained results are in agreement with numerical (Monte-Carlo) experiments.

In the above analysis we studied the model with zero hopping,
thus neglecting the quantum kinetic effects. In general, there may
be a drastic difference between the Hubbard models in the limits of
zero and small but nonzero hopping. This difference takes place for
models with spinful electrons and results from a huge spin degeneracy
of the ground state at zero hopping, while even a small hopping transfers
the half-filled state into an antiferromagnetic state (see, e.g., reviews \cite{Lee_Nagaosa} and
\cite{Aro}).  Fortunately, such a singularity is absent for the studied strongly disordered
model of spinless fermions where the weak hopping effects reduce
to a little smearing of the particle wave function localized at a given site \cite{Anderson1958}.
It seems plausible that the nonzero but weak hopping in such a situation will
not have a strong influence on the ``classical'' model.  Sure, accounting for a
nonzero hopping will be necessary in studies of dynamical (quantum) properties.

The studied model might have physical implementation in systems of 
cold atoms in optical lattices with randomly modulated on-site potentials.
Note that as far as we consider only two particles, the model is equally 
applicable also to hard-core bosons. In the article we have used the 
fermionic terminology having in mind future applications to 
many-particle systems. Formally, the model can also be mapped on the 
Ising spin chain in a random magnetic field but with an unnatural restriction 
to only two inverted spins sector of the Hilbert space.

The unveiled solvability of the one-dimensional \emph{disordered} model is a kind of serendipity.
It is not clear yet if there is a deeper mathematical reason behind the curtain.
At any rate, it would hardly make the things trivial: the transcendental equations
(\ref{eq:tan}) and (\ref{eq:tan_0}) for the eigenvalues do not look so.

It would be interesting to consider the generalization of the present theory 
to the case of an arbitrary (e.g., Gaussian) distribution of the site disorder $p(V)$,
when the transfer matrix is given by Eq.(\ref{eq:A-gen}) and the possibility of its 
analytical diagonalization is not \emph{a priori} obvious.

Finally, a generalization of the considered two-particle model (with an effective 
`filling factor' $2/N$) to a strongly disordered Hubbard-like model with a low 
but finite particle density is an appealing issue for future study.

\section{Acknowledgement}

We  thank G.V. Shlyapnikov for discussions at early stage of this work. 
V.Y. acknowledges the Basic research program of HSE.

	\vspace{2cm}

	\bibliographystyle{apsrev4-2}
	
%	\bibliography{bibliography}
	

\end{document}